\documentclass[11pt,letter,twocolumn]{article}

\usepackage{amssymb}
\usepackage{amsfonts}
\usepackage[fleqn]{amsmath}
\usepackage{graphicx}
\usepackage{bm}
\usepackage{caption}
\usepackage[pdftex,colorlinks,bookmarksopen,bookmarksnumbered,citecolor=blue,urlcolor=red]{hyperref}
\usepackage{subfigure}
\usepackage{xcolor}
\usepackage[numbers,comma,sort&compress]{natbib}
\title{\vspace{-4em}Towards understanding flexoelectricity at the nanoscale}
\author
{David~Codony$^{1,2,\ast}$, Phanish~Suryanarayana$^{1}$, Irene~Arias$^{3,2}$\\
\\
\small{$^1$ College of Engineering, Georgia Institute of Technology, Atlanta, GA 30332, USA,}
\\
\small{$^2$ Laboratori de C\`{a}lcul Num\`{e}ric (LaC\`{a}N), Universitat Polit\`{e}cnica de Catalunya, 08034 Barcelona, Spain}
\\
\small{$^3$ Centre Internacional de M{\`e}todes Num{\`e}rics en Enginyeria (CIMNE), 08034 Barcelona, Spain}
\\
\small{$^\ast$ Corresponding author; E-mail: dcodony@gatech.edu.} \\ {\small Article written for the 2023 ECCOMAS newsletter.} }
\date{}

\begin{document}
\maketitle

\noindent
\hrulefill

\vspace{-1em}
\noindent\hrulefill

\vspace{-1.5em}
{
\centering
\subsection*{Abstract}\noindent{\it
We review the authors' recent works on flexoelectricity at the nanoscale \cite{11,12}, while emphasizing the role of continuum mechanics in interpreting the electromechanical response of quantum mechanical systems under bending. }}

\vspace{-.5em}
\noindent\hrulefill

\vspace{-1em}
\noindent\hrulefill

\subsubsection*{Introduction}
\begin{figure}[h!]\centering  	\includegraphics[width=.63\linewidth,clip,keepaspectratio,angle=0]{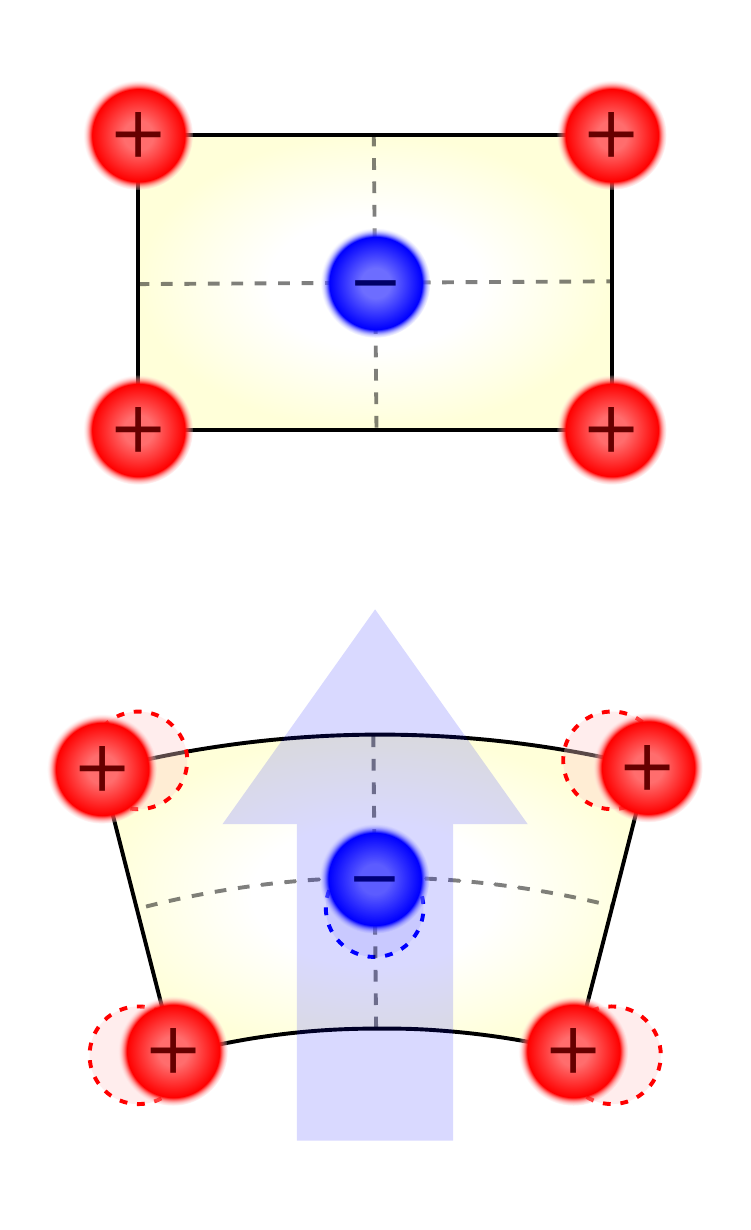}	
\caption{Ionic model for a crystal under bending deformations.} 
\label{fig:1}
\end{figure}

Flexoelectricity is a universal physical phenomenon that has its origins at the nanoscale. It represents an electromechanical two-way coupling between electric polarization and strain gradients, referred to as direct flexoelectricity, or between polarization gradients and strain, referred to as converse flexoelectricity. It is found to be significant in a wide range of materials, including cellular membranes, viruses, liquid crystals, polymers, ceramics, semiconductors, ferroelectrics, and atomic monolayers \cite{wang2019flexoelectricity}. In materials with a crystalline atomic structure, the flexoelectric effect has traditionally been conceptualized via the ionic crystal model under bending (Fig.~\ref{fig:1}), where the spatial inversion symmetry of the crystal is broken, resulting in a non-zero dipole moment, attributed to ionic flexoelectricity. In addition, there can be an electronic contribution to the dipole, attributed to electronic flexoelectricity. Flexoelectricity is not only restricted to bending, but can also be found in other non-uniform deformation settings such as torsion or inhomogeneous tension/compression. Unlike piezoelectricity, the flexoelectric effect is found in all dielectrics, regardless of their microscopic crystalline structure, becoming magnified at small length scales, where large strain gradients are commonly encountered.

The flexoelectric effect has a number of applications in new generation  electromechanical devices that do not rely on piezoelectric materials, circumventing the limitations typically associated with piezoelectric materials, which include brittleness, high lead content, and a narrow operative temperature range. This is particularly appealing for nanotechnological applications, including NEMS for nanoscale sensing and actuation, nanogenerators that harness energy from mechanical vibration, and ultra-high storage density memories  \cite{shu2019flexoelectric}. Other interesting applications at larger scales include the design of geometrically-polarized metamaterials with apparent piezoelectricity from non-piezoelectric materials, as well as flexible, biocompatible, and wearable energy harvesting devices \cite{shu2019flexoelectric}.

\subsubsection*{Continuum modeling}
\begin{figure*}[h!]\centering  	\includegraphics[width=.75\linewidth,clip,keepaspectratio,angle=0]{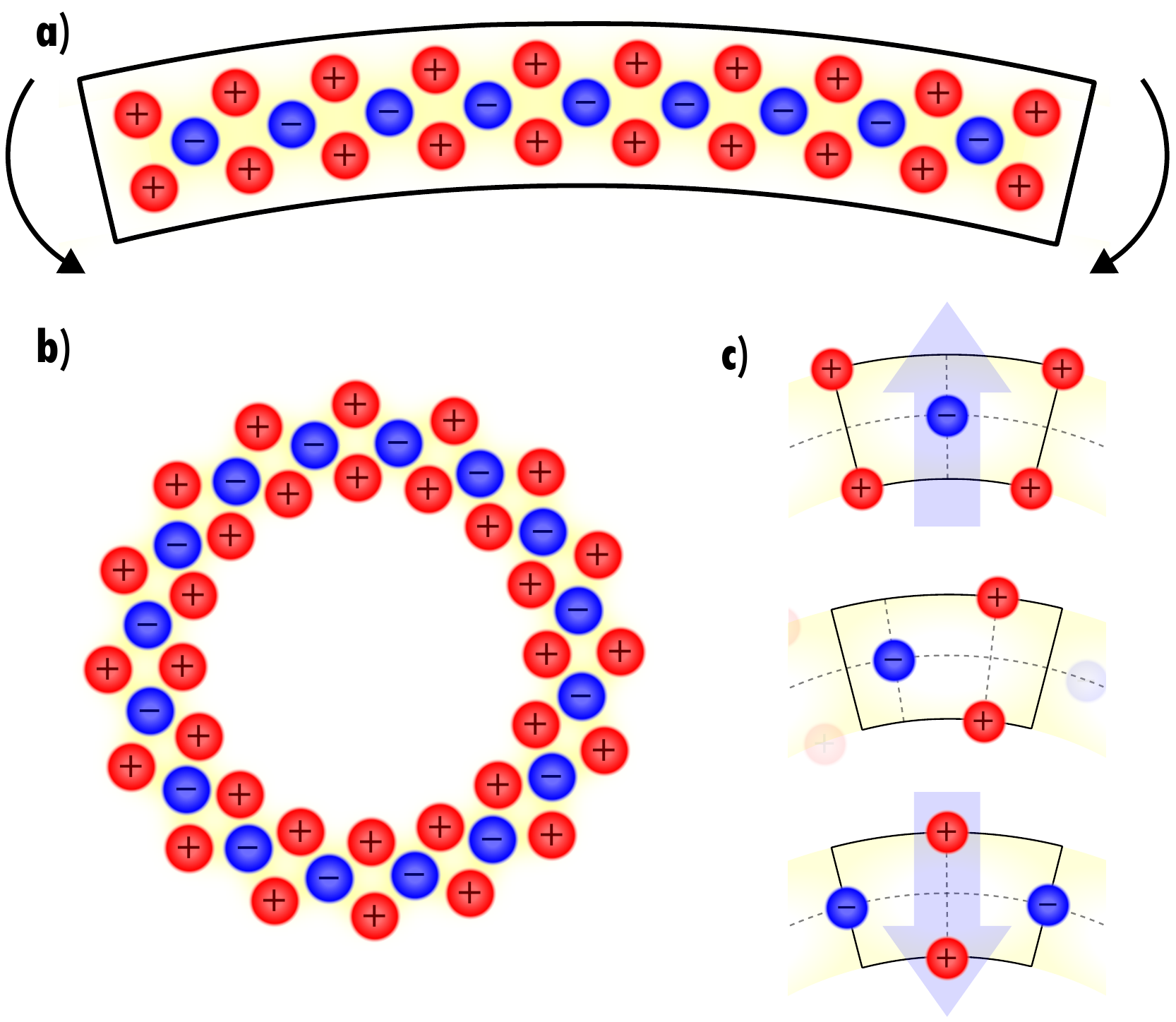}	
\caption{a) Finite 2D system under uniform bending, modeled as a continuum and as a finite quantum system. b) Quantum system bent until a cylindrical structure (nanotube) is formed. c) Different unit cells of the same system lead to different polarization states.} 
\label{fig:2}
\end{figure*}
Flexoelectricity can be modeled using \textbf{ continuum mechanics (CM)} \cite{06}. The most common continuum model for flexoelectricity in solids is inherited from strain gradient elasticity theory, which results in a fourth order coupled partial differential equation (PDE) for the displacement field and electric potential. When it comes to numerical calculation, standard finite element (FE) methods cannot be used due to the high-order nature of the PDE system, motivating the need for other computational frameworks that can deal with the $\mathcal{C}^1$-continuity requirement on the state variables. 

Besides the computational challenges, the modeling of flexoelectricity is associated with a number of open questions, assumptions and challenges. On the one hand, the direct and converse flexoelectric couplings — a combination of both is possible, known as Lifshitz-invariant coupling — lead to the same Euler-Lagrange equations but different admissible sets of boundary conditions, and thus different boundary value problems. In the computational flexoelectricity literature, usually only one of the many possible couplings is assumed, and it is not yet clear which boundary value problem for flexoelectricity is more appropriate.

On the other hand, at small length scales, there exist other physics/chemistry that may be relevant in certain situations. Boundary layers in the electric or mechanical fields have been reported in experiments, and are sometimes attributed to additional physical phenomena such as surface relaxation and/or surface piezoelectricity. The interplay between flexoelectric and piezoelectric effects must be considered in non-centrosymmetric materials, and it can be very counterintuitive given that piezoelectricity is not a size-dependent effect, contrary to flexoelectricity. Also, the Maxwell-stress effect plays a relevant role in a large deformation setting, which complicates analyses.

Another apparent difficulty lies in the constitutive laws not being fully validated by experimental evidence. Experiments to characterize the flexoelectric material constants entering the constitutive laws are relatively sparse, which can be attributed to the high cost of experiments, driven by the need for very high-resolution equipment. Furthermore, there can be up to orders of magnitude  disagreements in the reported values. Agreement with theoretical predictions or ab-initio calculations is also poor, sometimes even in the sign. Besides, the dependence of constitutive laws on temperature, magnetic fields, and/or other physics is generally overlooked, and a linear flexoelectric coupling is usually assumed.

\subsubsection*{Quantum mechanical modeling}
In order to study the flexoelectric effect at small length scales, an alternative is to use \textbf{quantum mechanics (QM)}, where matter is no longer assumed to be continuous, but rather treated a discrete collection of atoms (Fig.~\ref{fig:2}{\color{red}a}). Material behavior is then governed by Schrödinger’s equation, a linear PDE in the form of an eigenvalue problem, involving the so-called ``wavefunction'', a quantity that characterizes the quantum system. Simplified QM theories derived from Schrödinger’s equation, such as Kohn-Sham Density Functional Theory (DFT), are very popular, given their high accuracy to cost ratio. In this case, rather than solve for the many-body wavefunction, one essentially solves for the Kohn-Sham orbitals, which form the solution to a nonlinear eigenvalue problem. In so doing, the electron density of the system and relaxed nuclei positions can be determined. Indeed, once the ground-state has been determined, derived properties of interest can be calculated. This family of methods is devoid of constitutive laws or material parameters, and thus they are known as \emph{ab-initio} (from scratch) or first principles methods.

Computational DFT has its own limitations and challenges. On the one hand, the number of atoms in a system is a limiting factor: even the most sophisticated implementations can only handle simulations for systems up to a thousand atoms routinely. This means that the system cannot be larger than a few nanometers, i.e., smaller than the size of a virus. Indeed, systems with a well-defined crystal structure can be solved by exploiting their translational/periodic symmetry. However, in the case of flexoelectricity, the translational symmetry is broken by bending deformations, and/or other deformation modes that generate non-uniform strains. Computations accounting for a complete finite system (see Fig.~\ref{fig:2}{\color{red}a}) are relatively costly, and inevitably include undesired edge-related effects, which precludes the use of even state-of-the-art DFT codes \cite{07}.  To overcome this restriction, one can consider a material bent over onto itself until the two opposite edges merge, forming an extended cylindrical system that is free from edge effects (see Fig.~\ref{fig:2}{\color{red}b}), for which cyclic symmetry-adapted DFT \cite{08,09} can be used to perform efficient quantum mechanical calculations in a single unit cell, similar to the case of crystals. 

A major challenge in computational DFT is the \emph{interpretation} of the obtained results. One is typically not only interested in finding the electron density of a system or the associated energy, but also in extracting a macroscopic measurable. Examples for bent systems are the bending stiffness, or indeed the transversal flexoelectric coefficient, among others. For the latter, the flexoelectric coupling is typically modeled at a macroscopic level by a term in the energy of the system  that is linearly proportional to the strain gradient (curvature) and the electric polarization. Hence, by tracking the electric polarization of the system at different bent states, a fit can be made to obtain the transversal flexoelectric coefficient.

In the quantum mechanics DFT community, the electric polarization of a system can be formally identified with the volume-normalized dipole moment of the total charge, i.e., electron density $\rho(\mathbf{x})$ and the atom nuclei, represented as point charges $Z_i$. However, for a bent system, it remains unclear in which direction the polarization should be computed. For finite systems (Fig.~\ref{fig:2}{\color{red}a}), one could think of computing the polarization along the direction of the angle bisector; however, the result depends on the angle subtended by the system. For the extended structure  forming a cylinder (Fig.~\ref{fig:2}{\color{red}b}), the polarization computed along any transversal direction vanishes due to the symmetry of the system. If the polarization is computed in a single unit cell (Fig.~\ref{fig:2}{\color{red}c}), the result strongly depends on the choice of unit cell. This indicates that the polarization state in bent systems is  ill-defined.

\subsubsection*{Radial polarization formulation}

Continuum and quantum mechanics can be used together to properly understand, model, and compute flexoelectricity at the nanoscale.  The key observation is that the interpretation of polarization made in the QM community is flawed due to the consideration of an oversimplistic continuum model. By adopting a continuum framework for finite deformations, a uniform bending deformation map transforms a flat system into a curved one, turning its translational symmetry into a cyclic one. The careful consideration of these two different states, i.e., flat and bent, is crucial in this context.

On the one hand, the energy functional must be written in a Lagrangian frame for the flat system, so that frame invariance is fulfilled \cite{10}. Therefore, the associated flexoelectric tensor in this energy functional has a Lagrangian nature. On the other hand, quantum systems are described in a Eulerian frame, since the actual bent system is directly modeled without the need to know or describe the original flat state.

\begin{figure*}[h!]\centering  	\includegraphics[width=.8\linewidth,clip,keepaspectratio,angle=0]{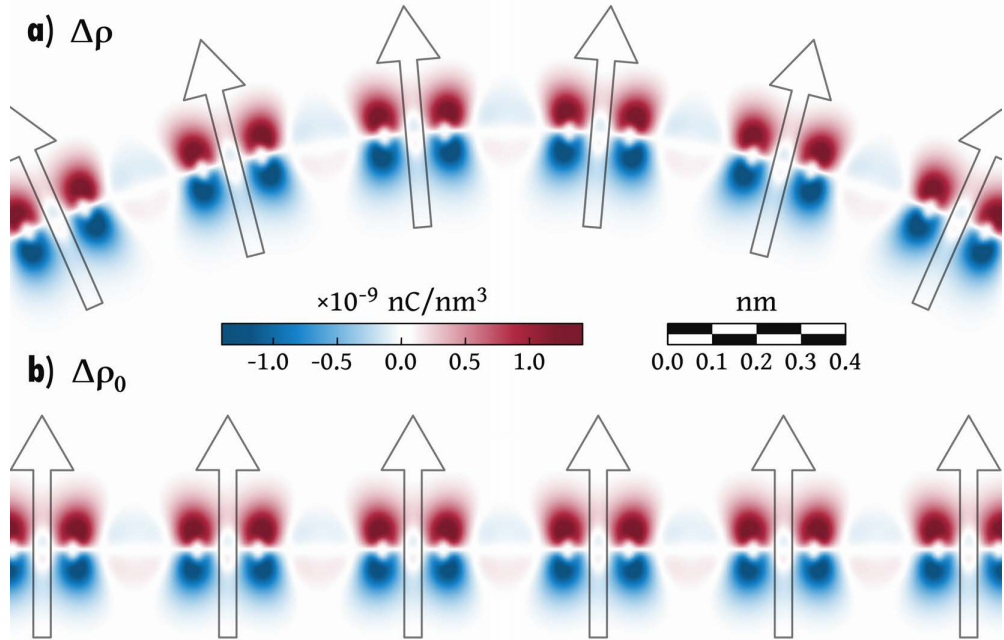}	
\caption{Electron charge density redistribution in graphene upon a curvature increase of $\Delta\kappa = 0.18$ 1/nm in a) Eulerian frame, and b) Lagrangian frame. The gray arrows indicate the direction along which the dipole must be computed in each frame.} 
\label{fig:3}
\end{figure*}

The characterization of the (Lagrangian) transversal flexoelectric coefficient in a quantum (Eulerian) system must be done carefully \cite{11}, as outlined next. Once the nuclei positions and electron density of the quantum system are found in the Eulerian frame (Fig.~\ref{fig:3}{\color{red}a}), they are pulled-back to the Lagrangian frame (Fig.~\ref{fig:3}{\color{red}b}), and the Lagrangian polarization is computed as the dipole moment of the nuclei and the Lagrangian electron density per unit volume ($\rho_0$), along the transversal direction. The transversal flexoelectric coefficient is then redefined as the rate of change of the Lagrangian polarization with respect to the curvature $\kappa$. This Lagrangian definition is independent of the choice of unit cell (Fig.~\ref{fig:2}{\color{red}c}) for extended systems (Fig.~\ref{fig:2}{\color{red}b}) and does not display an artificial dependence on the corresponding width for finite structures (Fig.~\ref{fig:2}{\color{red}a}), thereby overcoming a fundamental limitation of the standard definition. This demonstrates that the infinitesimal deformation assumption — widely used in the computational QM community — clearly leads to wrong results in the current context, and underpins the general need for sound continuum models in computational QM for the correct characterization of material properties. 

In order to work solely in a Eulerian frame, as typically done in computational QM, one can push-forward the definition of the Lagrangian polarization onto the Eulerian frame, which results in the dipole moment computed in the bent system along the \emph{radial} direction (indicated by gray arrows in Fig.~\ref{fig:3}{\color{red}a}). Such quantification for the polarization in a bent system has been coined as ``radial polarization'' $p_\textrm{r}$ \cite{11}, from which the transversal flexoelectric coefficient $\mu_\textrm{T}$ is computed as:
\begin{align*}
\mu_\textrm{T} &= \frac{\partial{p_\textrm{r}}}{\partial{\kappa}} \,,
\\
p_\textrm{r} &= \frac{1}{|\Omega|}\left(\sum_iZ_iR_i-\int_\Omega\rho(\mathbf{x})r\mathrm{d}\mathbf{x}\right) \,,
\end{align*}
where $r$ is the radial coordinate, $R_i$ is the radial position of the $i$-th nucleus, and $\Omega$ denotes the space occupied by the system. The $\kappa$-derivative of the radial polarization can be numerically approximated by computing $p_\textrm{r}$ at multiple curvatures $\kappa$ in the vicinity of the curvature at which $\mu_\textrm{T}$ is desired:
\begin{align*}
\mu_\textrm{T} &\approx \frac{\Delta{p_\textrm{r}}}{\Delta{\kappa}} \,,
\\
p_\textrm{r} &= \frac{1}{|\Omega|}\left(\sum_iZ_i\Delta R_i-\int_\Omega\Delta\rho(\mathbf{x})r\mathrm{d}\mathbf{x}\right) \,.
\end{align*}
It is worth noting that this formulation is not restricted to the linear regime, and can also be used to capture the nonlinear behavior. 

\begin{figure*}[htb!]\centering  	\includegraphics[width=\linewidth,clip,keepaspectratio,angle=0]{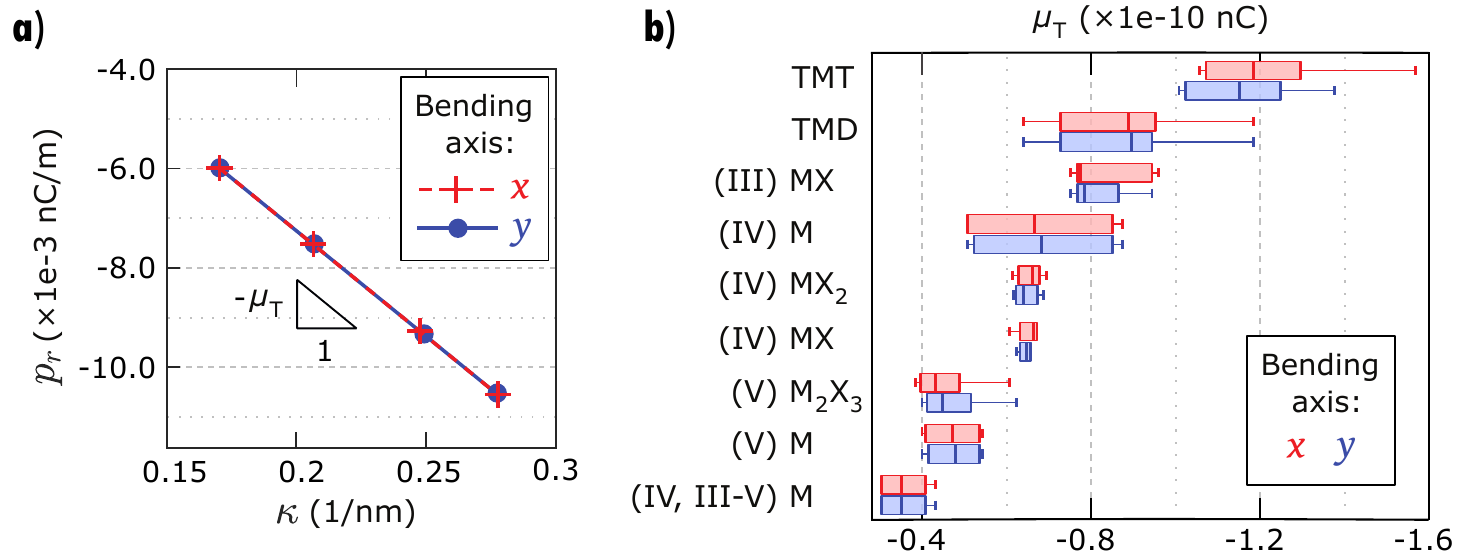}	
\caption{Calculation of the transversal flexoelectric coefficient in select atomic monolayers, data taken from Refs.~\cite{11,12}. a) Radial polarization in graphene bent at different curvatures. b) Box plot of the transversal flexoelectric coefficient for different monolayer families.} 
\label{fig:4}
\end{figure*}

\subsubsection*{Atomic monolayers}
Atomic monolayers are crystalline materials consisting of a single layer of atoms. They have gained in popularity since the synthesis of graphene in 2004. Generally, they present very exotic and unprecedented properties not found in their bulk counterparts. In the context of flexoelectricity, they are especially interesting due to their low bending-rigidity, which allows for large strain gradients and hence a large flexoelectric effect. In practical terms, it is convenient to first study nanoscale flexoelectricity in atomic monolayers, given their minimal thickness, which translates to small system sizes in computational QM simulations.  

As a demonstration of its capability, the radial polarization framework has been used to calculate the transversal flexoelectric coefficients for fifty-four atomic monolayers along their principal directions $(x,y)$ at 0 Kelvin \cite{12}. To calculate the radial polarization, the radial dipole is normalized with the area rather than the volume, due to the ill-defined nature of thickness in 2D-systems. Therefore, the flexoelectric coefficient values are reported in units of charge (Coulomb) rather than the usual charge per unit length (Coulomb/meter).

The following different families of 2D systems have been considered: Honeycomb lattice materials from Group III monochalcogenides (\texttt{III-MX}), Transition metal dichalcogenides (\texttt{TMD}), Groups IV, III-V and V monolayers (\texttt{M}), and Group IV dichalcogenides (\texttt{IV-MX$_2$}), as well as rectangular lattice materials from Group V monolayers (\texttt{V-M}), Group IV monochalcogenides (\texttt{IV-MX}), Transition metal trichalcogenides (\texttt{TMT}), and Group V chalcogenides (\texttt{V-M$_2$X$_3$}). Close to half of the selected systems have already been synthesized, with the rest also expected to be synthesized in the future, given that their stability has  been predicted by DFT.

The calculated transversal flexoelectric coefficient in all the systems is constant for the bending curvatures considered, which are commensurate with those found in experiments. This confirms that the flexoelectric coupling can be modeled as linear for the atomic monolayers in the regime studied. Fig.~\ref{fig:4}{\color{red}a} shows the bending-induced radial polarization in graphene bent at different curvatures. The slope of the linear fit ($R^2=1.000$) gives a transversal flexoelectric coefficient $\mu_\textrm{T}$ of $-0.35\cdot10^{-10}$ nC, indicating that graphene is one of the 2D materials in this study with the smallest flexoelectric response. On the other side of the spectrum, the ZrTe${}_3$ system has the largest response, with a value of $-1.57\cdot10^{-10}$ nC.

The characterization of $\mu_\textrm{T}$ for each of the monolayer families studied is summarized in Fig.~\ref{fig:4}{\color{red}b}. Insightfully, the flexoelectric coefficients are similar along both principal directions, irrespective of the lattice structure (honeycomb or rectangular). This suggests that the flexoelectric tensor in a continuum model for these atomic monolayers may be characterized by a transverse isotropic behavior.

The calculated flexoelectric coefficients agree well with available experimental data \cite{12}. Some of the systems even have a good quantitative agreement, e.g.~only 15\% difference in the case of the MoS${}_2$ system. The comparison to other DFT studies, which are also sparse and have been done for very few monolayers, is relatively poor. We attribute this significant difference mainly to the use of the ill-defined polarization in the Eulerian frame, considered in other works.

\subsubsection*{Future avenues of research}
The present review has highlighted the importance of incorporating proper continuum models to understand and characterize quantum mechanical systems, and in turn how can computational quantum mechanics validate the assumptions as well as provide information for continuum models, all in the context of nanoscale flexoelectricity for lower dimensional systems. The described framework is applicable to other 2D systems, including multilayered heterostructures and thin films — such as ferroelectric perovskites — where the flexoelectric response is expected to be magnified. There, boundary layer effects will likely arise and non-linear regimes may be found, the correct continuum modeling of which is a subject worthy studying. The framework can also be easily extended to 1D materials, which themselves demonstrate interesting and exotic properties. In so doing, deformation modes other than uniform bending, such as a combination of bending+torsion or non-uniform bending, can be considered. The dependence of the obtained results with temperature can also be studied via molecular dynamics (MD) simulations.

In addition to flexoelectricity, other interesting material properties can be formulated within the described framework, including the flexophotovoltaic effect, flexomagnetism, and the change of carrier mobility with curvature in semiconductive systems under bending.

\subsubsection*{Acknowledgements}
D.C.~acknowledges the support of the Spanish Ministry of Universities through the Margarita Salas fellowship (European Union - NextGenerationEU). P.S.~acknowledges the U.S.~National Science Foundation and the Clifford and William Greene, Jr.~Professorship. I.A.~acknowledges the support of the European Research Council (StG-679451), Generalitat de Catalunya (2017-SGR-1278 and the ICREA Academia award) and the Spanish Ministry of Economy and Competitiveness through the Severo Ochoa Programme for Centres of Excellence in R\&D (CEX2018-000797-S).

\bibliography{ref}

\end{document}